\documentclass[11pt,twoside]{article}

\usepackage{asp2014}

\aspSuppressVolSlug
\resetcounters

\begin{document}

\title{Open-source Analysis Tools for Multi-instrument Dark Matter Searches}

\author{T. Miener$^1$, D. Kerszberg$^2$, C. Nigro$^2$, J. Rico$^2$, and D. Nieto$^1$}
\affil{$^1$Instituto de F\'{i}sica de Part\'{i}culas y del Cosmos and Departamento de EMFTEL, Universidad Complutense de Madrid, E-28040 Madrid, Spain; \email{tmiener@ucm.es}}
\affil{$^2$Institut de F\'{i}ısica d’Altes Energies (IFAE), The Barcelona Institute of Science and Technology (BIST), E-08193 Bellaterra (Barcelona), Spain}

\paperauthor{T. Miener}{tmiener@ucm.es}{https://orcid.org/0000-0003-1821-7964}{Universidad Complutense de Madrid, Madrid, Spain}{Instituto de F\'{i}sica de Part\'{i}culas y del Cosmos and Departamento de EMFTEL}{Madrid}{Madrid}{E-28040}{Spain}
\paperauthor{D. Kerszberg}{dkerszberg@ifae.es}{https://orcid.org/0000-0002-5289-1509}{Institut de F\'{i}ısica d’Altes Energies (IFAE)}{The Barcelona Institute of Science and Technology (BIST)}{Bellaterra (Barcelona)}{Barcelona}{E-08193}{Spain}
\paperauthor{C. Nigro}{cosimo.nigro@ifae.es}{https://orcid.org/0000-0001-8375-1907}{Institut de F\'{i}ısica d’Altes Energies (IFAE)}{The Barcelona Institute of Science and Technology (BIST)}{Bellaterra (Barcelona)}{Barcelona}{E-08193}{Spain}
\paperauthor{J. Rico}{jrico@ifae.es}{https://orcid.org/0000-0003-4137-1134}{Institut de F\'{i}ısica d’Altes Energies (IFAE)}{The Barcelona Institute of Science and Technology (BIST)}{Bellaterra (Barcelona)}{Barcelona}{E-08193}{Spain}
\paperauthor{D. Nieto}{d.nieto@ucm.es}{https://orcid.org/0000-0003-3343-0755}{Universidad Complutense de Madrid, Madrid, Spain}{Instituto de F\'{i}sica de Part\'{i}culas y del Cosmos and Departamento de EMFTEL}{Madrid}{Madrid}{E-28040}{Spain}

\begin{abstract}
The nature of dark matter (DM) is still an open question in Physics. Gamma-ray and neutrino telescopes have been searching for DM signatures for several years and no detection has been obtained so far. In their quest, these telescopes have gathered a wealth of observations that, if properly combined and analyzed, can improve on the constraints to the nature of DM set by individual instruments. In this contribution, we present two open-source analysis tools aimed at performing the before mentioned combined analysis: \texttt{gLike}, a general-purpose ROOT-based code framework for the numerical maximization of joint likelihood functions, and \texttt{LklCom}, a Python-based tool combining likelihoods from different instruments to produce combined exclusion limits on the DM annihilation cross-section.
\end{abstract}

\section{Introduction}

In the context of open-source tools and multi-messenger astronomy, it is essential to provide the community with reliable and accessible analysis tools to reproduce state-of-the-art scientific results and to facilitate the incorporation of those tools in custom analysis chains for future work. We present two open-source code frameworks that combine multi-instrument data from gamma-ray and neutrino telescopes and that perform a global analysis aimed to constrain the nature of DM.

\newpage
\section{Analysis workflow}

The workflow for the \texttt{gLike}\footnote{\href{https://github.com/javierrico/gLike}{https://github.com/javierrico/gLike}}~\citep{javier_rico_2021_4601451} and \texttt{LklCom}\footnote{\href{https://github.com/TjarkMiener/likelihood_combiner}{https://github.com/TjarkMiener/likelihood\_combiner}}~\citep{tjark_miener_2021_4597500} tools is depicted in Fig~\ref{fig:UMLWorkflow}. \texttt{gLike} is a self-containing framework, which takes as input high-level gamma-ray astronomical data in a format interfaceable with the one proposed by the \texttt{GADF}\footnote{\href{https://gamma-astro-data-formats.readthedocs.io/}{https://gamma-astro-data-formats.readthedocs.io/}} initiative ~\citep{deil_christoph_2018_1409831} and numerically maximizes the joint likelihood functions. The obtained likelihood curves as a function of the DM self-annihilation cross-section $ \langle\sigma v\rangle $ can be stored in an intermediate data format in txt and be later reutilized as input. The \texttt{LklCom} is fed with either the former curves stored as txt files or with a single hdf5 file, obtained by merging the txt files via its i/o tools. Various gamma-ray and neutrino telescopes can adopt the data format for the likelihood curves in their analysis chains and use the two presented analysis tools to perform a combined analysis. The tremendous advantage of this procedure is that there is no need of sharing sensitive low-level data among the participating instruments. \texttt{LklCom} features several matplotlib-based plotting functions to visualize the final output products generated by either tool.

\articlefigure[width=1.0\textwidth]{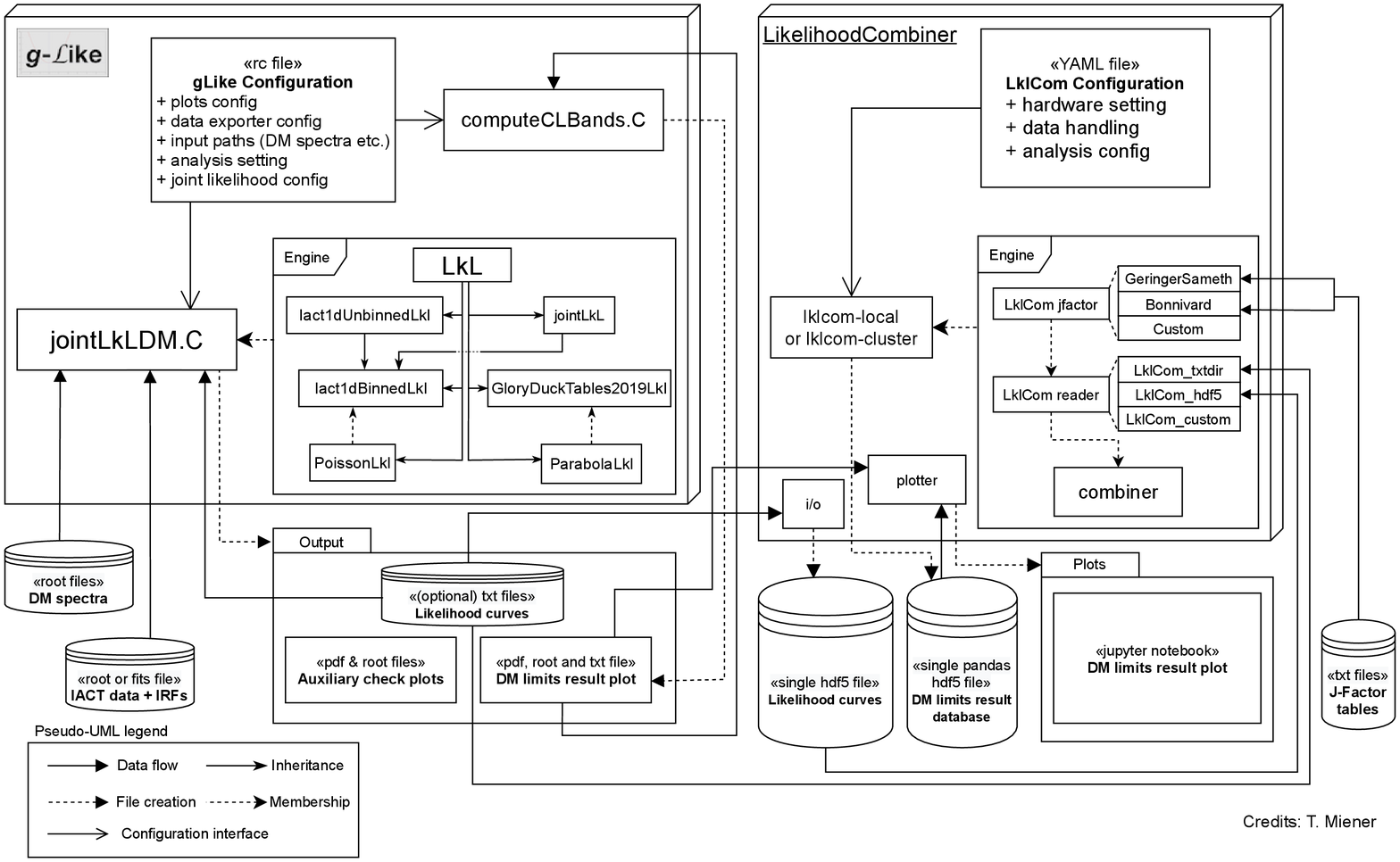}{fig:UMLWorkflow}{Pseudo-UML workflow from \texttt{gLike} and \texttt{LklCom}.}

\section{Results}

In order to verify the agreement of the independently implemented code frameworks, mock data is used and analyze with the same analysis setting. In Fig.~\ref{fig:Comparison}, we overlay the \texttt{LklCom} DM limits results plot with the DM limits and error bands obtained with \texttt{gLike} as red dashed contour lines, which shows an accurate agreement.

\articlefigure[width=0.6\textwidth]{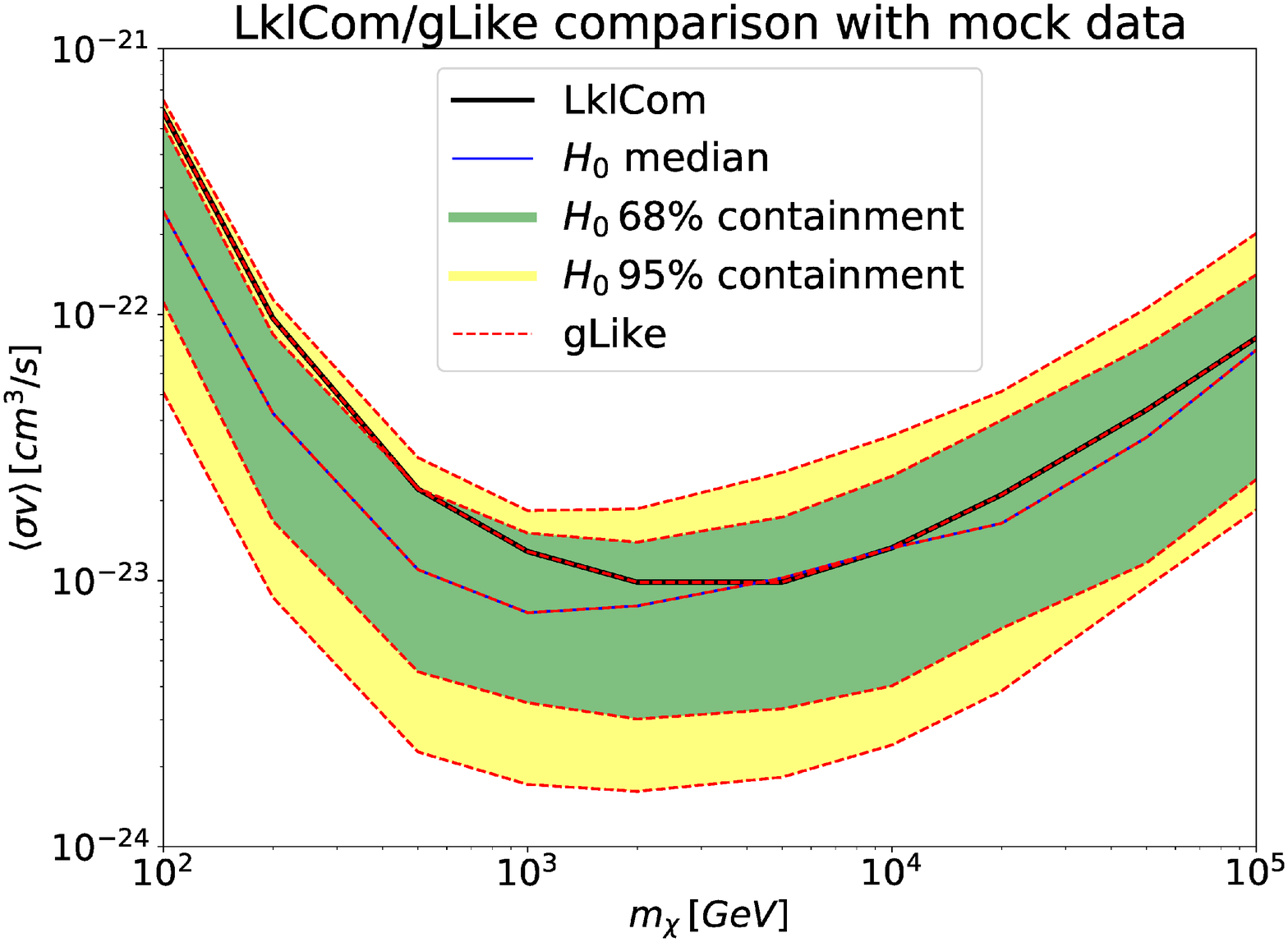}{fig:Comparison}{Comparison of \texttt{gLike} and \texttt{LklCom} with mock data.}

\noindent
The latest scientific results (see Fig.~\ref{fig:Armand}) of a global indirect DM search in the gamma-ray band, carried out with the presented tools, is ~\citep{Armand:2021},  where the major gamma-ray observatories \textit{Fermi}-LAT, HAWC, H.E.S.S., MAGIC, and VERITAS jointly analyzed observations of 20 dwarf spheroidal galaxies (dSphs). This work sets the most constraining and robust upper limits of the weakly interacting massive particles DM self-annihilation cross-section $ \langle\sigma v\rangle $ towards dSphs over the widest DM mass range, extending from 5 GeV to 100 TeV.

\articlefigure[width=1.0\textwidth]{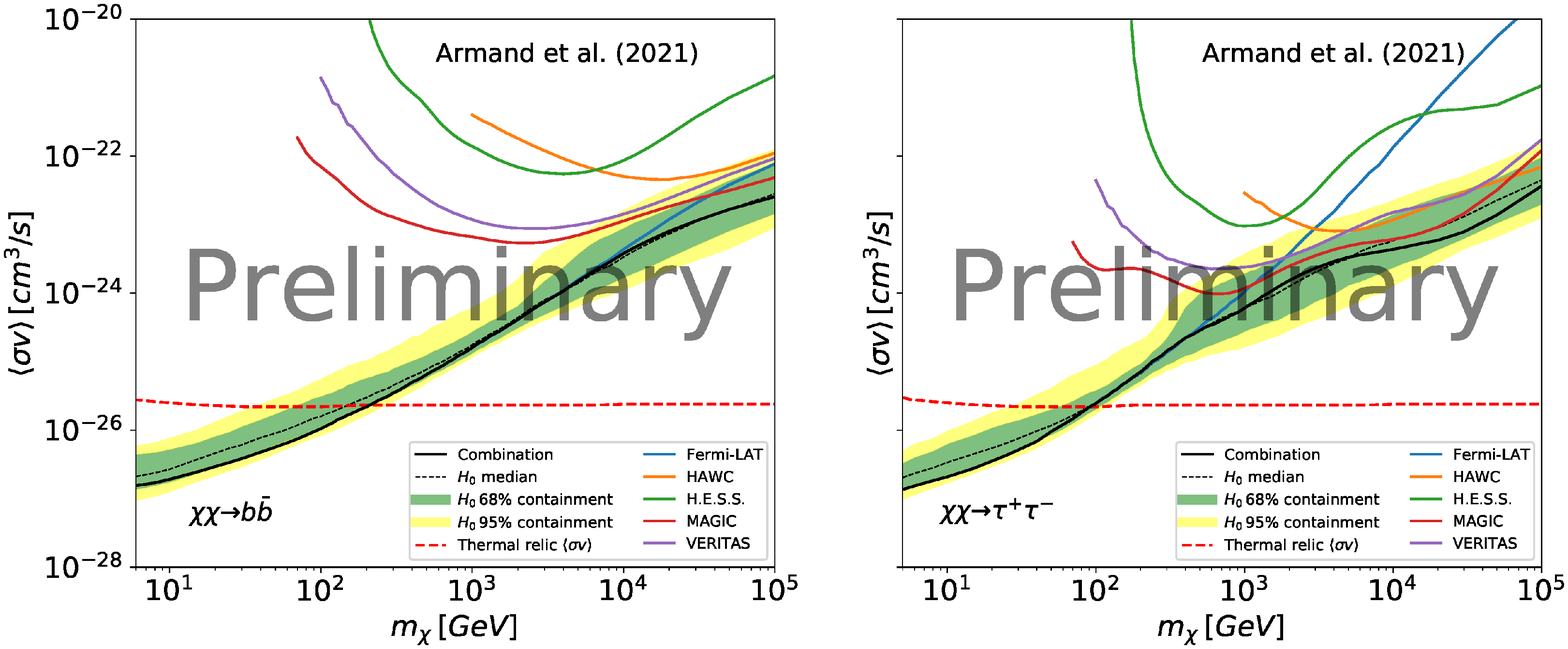}{fig:Armand}{[Taken from ~\citep{Armand:2021}] Upper limits at 95\% confidence level on $ \langle\sigma v\rangle $ as a function of the DM mass for the annihilation channels $ b\bar{b} $ (left) and $ \tau^{+}\tau^{-} $ (right) from the combined analysis of dSph observations by \textit{Fermi}-LAT, HAWC, H.E.S.S., MAGIC, and VERITAS performed by \texttt{gLike} and \texttt{LklCom}.}

\section{Conclusions and outlook}

This contribution presents two open-source software tools aiming to help the astronomical community to unify the search for DM and derive global DM constraints from multi-instrument and multi-messenger observations.

\acknowledgements
TM acknowledges support from PID2019-104114RB-C32. CN, JR, and DN acknowledges partial support from The European Science Cluster of Astronomy \& Particle Physics ESFRI Research Infrastructures funded by the European Union’s Horizon 2020 research and innovation program under Grant Agreement no. 824064. This work had support from the ERDF under the Spanish Ministerio de Ciencia e Innovaci\'{o}n (MICINN, grant PID2019-107847RB-C41), and from the CERCA program of the Generalitat de Catalunya.
\\
\\

\bibliography{X0-006}

\end{document}